\def\asca{{\it ASCA\/}}

\def\einstein{{\it Einstein\/}}

\def\hst{{\it {\it HST}\/}}
\def\hut{{\it {\it HUT}\/}}

\def\iue{{\it IUE\/}}
\def\rosat{{\it ROSAT\/}}

\def\aox{{$\alpha_{\rm ox}$}}

\def\ltsima{$\; \buildrel < \over \sim \;$}
\def\simlt{\lower.5ex\hbox{\ltsima}}
\def\gtsima{$\; \buildrel > \over \sim \;$}
\def\simgt{\lower.5ex\hbox{\gtsima}}
        
\documentstyle[12pt,psfig,aaspp4]{article}

\begin{document}


\title{On the Nature of Soft X-ray Weak Quasi-Stellar Objects}


\author{W. N. Brandt}
\affil{Department of Astronomy and Astrophysics,
525 Davey Laboratory,
Pennsylvania State University,
University Park, PA 16802}

\author{A. Laor}
\affil{Physics Department,
Technion,
Haifa 32000 Israel}

\author{Beverley J. Wills}
\affil{Department of Astronomy, 
University of Texas at Austin,
Austin, TX 78712}


\begin{abstract}
Recent studies of Quasi-Stellar Objects (QSOs) with \rosat\ suggest the 
existence of a significant population of Soft X-ray Weak QSOs (SXW~QSOs) 
where the soft X-ray flux is $\sim$~10--30 times smaller than in typical 
QSOs. Why are these QSOs soft X-ray weak, and what is their relationship 
to Broad Absorption Line QSOs (BAL~QSOs) and X-ray warm absorber QSOs?
As a first step in a systematic study of these objects, we 
establish a well-defined sample of SXW~QSOs which includes 
all $\alpha_{\rm ox}\leq -2$ QSOs from the Boroson \& Green (1992) 
sample of 87 Bright Quasar Survey QSOs. 
SXW~QSOs comprise $\approx 11$\% of this optically selected QSO sample, 
and we find soft X-ray weakness in both radio-quiet and radio-loud QSOs. 
From an analysis of C~{\sc iv} absorption 
in the 55 BG92 QSOs with available C~{\sc iv} data, we 
find a remarkably strong correlation between 
$\alpha_{\rm ox}$ and the C~{\sc iv} absorption equivalent
width. This correlation suggests that
absorption is the primary cause of soft X-ray weakness in QSOs, 
and it reveals a continuum of absorption properties connecting
unabsorbed QSOs, X-ray warm absorber QSOs, SXW~QSOs and BAL~QSOs.
Many of our SXW~QSOs have ultraviolet absorption that is intermediate
in strength between that of X-ray warm absorber QSOs and that of 
BAL~QSOs, and their X-ray absorption is also likely to be of
intermediate strength. From a practical point of view, our
correlation demonstrates that selection by soft X-ray weakness is 
an effective ($\simgt 80$\% successful) and observationally inexpensive 
way to find low-redshift QSOs with strong and interesting ultraviolet 
absorption. 

We have also identified several notable differences between the optical
emission-line properties of SXW~QSOs and those of the other 
Boroson \& Green QSOs. SXW~QSOs show systematically low [O~{\sc iii}] 
luminosities and equivalent widths as well as 
distinctive H$\beta$-line profiles. They tend to 
lie toward the weak-[O~{\sc iii}] end of Boroson \& Green eigenvector~1, 
as do many low-ionization BAL~QSOs. Unabsorbed Seyferts and QSOs with 
similar values of eigenvector~1 have been suggested to have 
extreme values of a primary physical parameter, perhaps mass accretion 
rate relative to the Eddington rate ($\dot M/\dot M_{\rm Edd}$). 
If these suggestions are correct, it is likely that SXW~QSOs also
tend to have generally high values of $\dot M/\dot M_{\rm Edd}$. 
Finally, we present and discuss correlations between \aox\ and
other QSO observables after removal of the SXW~QSOs. 

\end{abstract}


\keywords{
galaxies: active~--
galaxies: nuclei~--
quasars: general~--
X-rays: galaxies.
}


\section{Introduction}

While X-ray emission appears to be a universal property of Quasi-Stellar 
Objects (QSOs; Avni \& Tananbaum 1986, hereafter AT86), there is evidence 
for a significant population of soft X-ray weak QSOs (SXW~QSOs; 
Elvis 1992; Laor et~al. 1997a; Yuan et~al. 1998).
These comprise $\sim 10$\% of the optically selected
QSO population and have soft X-ray emission that is $\simgt$10--30 times 
weaker than expected based on their continuum luminosities in other bands. 
In general, the cause of soft X-ray weakness has not  
been well established, and it could arise as a result of 
(1) intrinsic or intervening X-ray absorption,
(2) an unusual underlying spectral energy distribution, or 
(3) extreme X-ray or optical variability. 

In this paper, we investigate the X-ray, ultraviolet
and optical properties of a well-defined sample of SXW~QSOs. 
Our main goals are
(1) to determine if soft X-ray weakness correlates with other observables, 
(2) to examine the relationship between SXW~QSOs, Broad Absorption Line QSOs (BAL~QSOs)
and QSOs with X-ray warm absorbers, and 
(3) to determine the reason for the low apparent soft X-ray luminosities of SXW~QSOs. 
Regarding the second goal, it is known that BAL~QSOs are 
generally weak in the soft X-ray band 
(e.g., Green \& Mathur 1996), but the fraction
of SXW~QSOs that are BAL~QSOs is poorly determined at present. 
Our focus is on objects that are more luminous and distant than 
local Seyferts (e.g., Maiolino \& Rieke 1995). 
Our work differs from that of Green (1998) in that we study
a smaller number of objects that are substantially more extreme 
in terms of their soft X-ray weakness. 
 

\section{Soft X-ray Weak QSO Selection}

We have selected our SXW~QSOs from the Boroson \& Green (1992; hereafter 
BG92) sample. This sample is composed of all 87 QSOs from the Bright Quasar 
Survey (BQS; Schmidt \& Green 1983) with $z<0.5$, and our intention is
to perform a thorough analysis of all the SXW~QSOs in this sample. 
The BQS consists of the broad-line active galaxies
with dominant star-like nuclei to the survey limits: $B\simlt 16.2$ 
and $U-B\simlt -0.44$ (for Galactic latitudes greater than $30^\circ$ and 
declinations above $-10^\circ$). To these limits it is thought to be 
$\simgt 70$\% complete at $z<0.5$ (see V\'eron et~al. 1999), 
and a large amount of high-quality and uniformly analyzed 
data are available for the BG92 QSOs. Since the 
BQS~QSOs were optically selected, the BQS should not be {\it directly\/} 
biased with respect to the inclusion or exclusion of SXW~QSOs. Possible 
{\it indirect\/} biases are, unfortunately, inescapable when using 
current large-area QSO surveys (see Goodrich 1997 and Krolik \& Voit 1998 
for examples of potentially relevant indirect biases). 

Our selection method is based upon \aox, the slope of a
power law defined by the rest-frame flux densities at 
3000~\AA\ and 2~keV. That is, 
$\alpha_{\rm ox}=0.372 \log(f_{\rm 2~keV}/f_{\rm 3000~\AA})$
where $f_{\rm 2~keV}$ and $f_{\rm 3000~\AA}$ are flux
densities.\footnote{Some authors prefer
to work with $\alpha^\prime_{\rm ox}$, the slope of a
power law defined by the rest-frame flux densities at
2500~\AA\ and 2~keV. For a 2500~\AA\ to 3000~\AA\ 
slope of $\alpha_{\rm u}$, we find 
$\alpha^\prime_{\rm ox}=1.03\alpha_{\rm ox}-0.03\alpha_{\rm u}$.
Note that $\alpha^\prime_{\rm ox}$, $\alpha_{\rm ox}$ and
$\alpha_{\rm u}$ are negative numbers.} 
While \aox\ calculations for many of the 
BQS~QSOs have been performed before (e.g., Tananbaum et~al. 1986), these
are not sufficient for our purposes as they miss many of the SXW~QSOs 
that we study here. These objects were either not observed or 
not detected. In addition, comparison of our \aox\ values with 
earlier ones allows us to search for \aox\ variability; extreme X-ray
or optical variability could cause a QSO to appear soft X-ray weak
(e.g., see Figure~1 of Boller et~al. 1997). We have therefore calculated 
our own \aox\ values for the 87 BG92 QSOs following the method 
in the next paragraph. 

When possible we have determined the 3000~\AA\ flux densities 
needed for \aox\ using the 
data from Neugebauer et~al. (1987). For the 13 BG92 QSOs 
without such data, we calculated 3000~\AA\ flux densities 
using data from
De Bruyn \& Sargent (1978; 4 QSOs),
Laor et~al. (1997ab; 2 QSOs), 
Neugebauer et~al. (1979; 1 QSO) and 
Schmidt \& Green (1983; 6 QSOs). 
We have used \rosat\ data to determine 
the requisite 2-keV flux densities
of our QSOs. We obtained count rates from the \rosat\ All-Sky 
Survey (RASS; Voges et~al. 1999) and pointed
observations. All but four of our objects (0043+039, 1004+130, 
1259+593, 1700+518) have \rosat\ detections. Of these four,
we note that 0043+039, 1004+130 and 1700+518 were also 
undetected in moderately deep \asca\ and \einstein\ 
observations (Elvis \& Fabbiano 1984; Gallagher et~al. 1999).
We converted the count rates into 2-keV flux densities using 
a power-law model with Galactic absorption. 
We used the Galactic column densities from 
Elvis, Lockman \& Wilkes (1989), 
Lockman \& Savage (1995) and 
Murphy et al. (1996). 
The power-law energy 
index ($\alpha_{\rm x}$) was predicted based on a fit to the 
H$\beta$~FWHM/$\alpha_{\rm x}$ correlation shown in 
Figure~5a of Laor et~al. (1997a):
$\alpha_{\rm x}=6.122-1.277~\log({\rm H\beta~FWHM})$.
The H$\beta$~FWHM/$\alpha_{\rm x}$ correlation has been 
directly established to hold for $\approx 1/4$ of the QSOs in 
our sample, and our fit is valid for H$\beta$~FWHM in the
range $\approx$~800--11000~km~s$^{-1}$. We excluded
3C~273 from our fit to prevent it from skewing 
the result due to its small error bar. 

The resulting \aox\ values are given in Table~1, and we show a histogram 
of these \aox\ values in Figure~1. For 0043+039 and 1004+130, we were 
able to place substantially tighter constraints upon \aox\ using 
the \asca\ and \einstein\ results (see above). The \aox\ distribution 
is non-Gaussian with $>99$\% confidence due to its tail towards large 
negative values of \aox. 
We found acceptable agreement between our \aox\ values and those 
for the 23 QSOs given in Table~4 of Laor et~al. (1997a). The 
average absolute difference in \aox\ for these 23 QSOs is 0.060, 
and no systematic discrepancies are apparent. We expect 
any systematic errors to be smaller than the random errors 
that are inherently present due to the X-ray and optical 
variability of QSOs. 
Furthermore, we have verified that none of the main results
below depends upon the precise details of how we calculate \aox. 
For example, we recover the same main results if we use
$\alpha_{\rm x}=1.72$ (1.15) for all radio-quiet QSOs 
(radio-loud QSOs) in the \aox\ calculation above 
(see \S4.1 of Laor et~al. 1997a). 

In this paper we have defined a SXW~QSO to be a QSO with 
$\alpha_{\rm ox}\leq -2$. These SXW~QSOs thus have 2-keV flux
densities that are $\simgt 25$ times weaker than expected
for a `normal' QSO (see \S4.3 of Laor et~al. 1997a). While 
the precise choice of the \aox\ division value is somewhat 
arbitrary, inspection of the histogram shown in Figure~1 
indicates that our choice is a reasonable one. The objects 
with $\alpha_{\rm ox}\leq -2$ stand out clearly from 
the bulk of the QSO population.
We have tested for bimodality of our \aox\ 
distribution using the {\sc kmm} mixture-modeling algorithm 
described by Ashman, Bird \& Zepf (1994, hereafter ABZ94). This 
algorithm indicates bimodality of the distribution
with high statistical significance
(the derived {\sc kmm} $P$-value is $<0.02$; 
see \S2.1 of ABZ94). The algorithm places the 10 QSOs 
with the largest negative values of \aox\ into one group 
(corresponding to $\alpha_{\rm ox}\leq -2$) and the
rest of the QSOs into another. The only QSO with an 
ambiguous {\sc kmm} classification is 2251+113; this object
is discussed further in \S6. However, since the {\sc kmm} 
algorithm makes several statistical assumptions 
(e.g., Gaussian-distributed populations; see Hartigan \& Hartigan 1985
and ABZ94), we regard our bimodality result as 
suggestive rather than conclusive. With a sample of 87 QSOs,
a distribution with a significant skew tail towards large 
negative \aox\ (compare \S2 of AT86) could mimic a bimodal one. 
A nonparametric Hartigan \& Hartigan (1985) dip test is
unable to rigorously establish bimodality, but this certainly 
does not rule out bimodality since the dip test sacrifices a great
deal of statistical power for its robustness. 
The histogram in Figure~1 may be compared with Figure~1 of 
AT86, although we note that our \aox\ values are substantially 
more complete and constraining. Our histogram may also be compared
with Figure~10 of Yuan et~al. (1998), although in this case
our sample is more statistically complete and homogeneous
(see \S2 of AT86 for a discussion of the importance of complete
samples).

We find 10 SXW~QSOs among the 87 QSOs in the BG92 sample, and 
these are listed in Table~2. Thus, SXW~QSOs comprise
$\approx 11$\% of the optically selected QSO population. This 
percentage is in reasonable agreement with the `detections and
bounds' upper limit of AT86.
All of our SXW~QSOs are radio quiet (with $\log R<1$)
except for 1004+130.\footnote{Here and hereafter we take 
$R$ to be the ratio of the observed 
6~cm and 4400~\AA\ flux densities.}
In Table~2 we have also given the historical values for \aox\ 
from \einstein\ observations (Tananbaum et~al. 1986); 
these values will be used in \S8. 
It is possible that 1259+593 is a SXW~QSO as well, but 
unfortunately the current X-ray data do not allow a tight 
upper limit to be placed upon \aox\ for this object 
($\alpha_{\rm ox}<-1.75$). 
We also comment that 0844+349 was seen to be fairly weak in soft 
X-rays during a \rosat\ pointed observation but not during the 
RASS (Yuan et~al. 1998). Here we have adopted the RASS count 
rate, and we discuss this object further in \S6. 


\section{Statistical Analyses of Continuum and Optical Emission-Line Properties}

We have performed statistical tests to compare the continuum and 
optical emission-line properties of SXW~QSOs with those 
of the bulk of the QSO population. In particular, we have compared 
the properties of our 10 SXW~QSOs (hereafter SXW for this group) 
to those of the other 77 BG92 QSOs (hereafter non-SXW for this group)
using two-sample Kolmogorov-Smirnov (hereafter KS) and Kuiper tests 
(see Press et al. 1992). We chose to apply both of these tests because
they statistically complement each other. These tests were 
objectively chosen in an a priori manner; we did {\it not\/} 
inspect the data in detail and then decide which statistical 
tests to apply. 
The main properties we compare are those listed 
in Tables~1 and 2 of BG92.\footnote{For 1211+143 we 
have corrected the $R$ parameter following Kellermann et~al. (1994). 
We have also corrected the H$\beta$~FWHM of 1307+085
to be 5320~km~s$^{-1}$ and the H$\beta$~shift of 1351+236 to be 1.099.} 
In addition, we compare 
the degree of broad-band optical continuum polarization 
(3200--8600~\AA; given in Table~1 of Berriman et~al. 1990),
the optical continuum slope ($\alpha_{\rm opt}$; given in 
Table~5 of Neugebauer et~al. 1987), and
the radio structure (e.g., Miller, Rawlings \& Saunders 1993;
Kellermann et~al. 1994). 
The probabilities resulting from our KS and Kuiper tests are listed 
in Table~3, and some of the continuum and optical emission-line 
properties are illustrated in Figure~2.

\subsection{Similarities}

In terms of many of their properties, SXW~QSOs are homogeneously 
distributed among the QSO population. For example, there are no 
statistically significant differences between SXW and non-SXW for 
$M_V$, $z$, or $R$. Of the 71 BG92 radio-quiet QSOs, 9 
are soft X-ray weak. Of the 16 BG92 radio-loud QSOs, 1 
is soft X-ray weak (the radio-loud QSO 
2251+113 is also quite weak in the soft X-ray band; see \S6). 
These rates of occurrence are statistically consistent according to 
a Fisher exact probability test for a $2\times 2$ contingency table. 
In addition, the equivalent width (EW) distributions of 
H$\beta$, He~{\sc ii} and Fe~{\sc ii} are consistent between 
SXW and non-SXW. 

As a class, our SXW~QSOs do not show higher broad-band optical 
continuum polarization than the non-SXW QSOs. Of 
our 10 SXW~QSOs, 9 have broad-band polarization 
percentages smaller than 0.9\%. The one notable exception is 1535+547 
which showed the highest polarization percentage in the BQS: 2.5\% 
(see Berriman et~al. 1990). We give additional notes on the polarization 
properties of our SXW~QSOs in \S5. 
We also find no evidence for a statistically significant difference 
between the optical continuum slopes of SXW and non-SXW
(i.e. the SXW~QSOs do not appear to be systematically redder
than the other BG92 QSOs). Although we note that both SXW and 
non-SXW were drawn from the BQS which employed selection by 
ultraviolet excess, the range in ultraviolet excess for the BQS 
is sufficiently large that a real difference would not necessarily 
be masked. 

\subsection{Differences}

There are several ways in which the optical emission-line 
properties of SXW~QSOs stand out from those of the bulk of 
the optically selected QSO population. 
The most notable one is that SXW~QSOs systematically have low 
[O~{\sc iii}] luminosities, and we have further illustrated this
in Figure~3. 
In addition, while the H$\beta$-FWHM and H$\beta$-asymmetry 
distributions of SXW and non-SXW are consistent, they have 
significantly different H$\beta$-shift and H$\beta$-shape 
distributions. 
The H$\beta$-shape distributions of SXW and non-SXW are found 
to be inconsistent by both the KS and Kuiper tests, and SXW~QSOs 
preferentially have `peaky' line profiles with concave sides. 
The H$\beta$-shift distributions of SXW and non-SXW 
are only shown to be different with high statistical significance 
by the Kuiper test. This result can be understood by inspection 
of \S14.3 of Press et~al. (1992) and Figure~2. The SXW~QSOs tend 
to lie at either the redward or blueward extremes of the H$\beta$-shift 
parameter, and this is where the KS test is least sensitive but 
the Kuiper test is significantly more sensitive (i.e. the KS
test is good at finding differences in the median value but not 
as good at finding differences in the `spread'). While we must 
recommend some caution regarding the H$\beta$-shift result
since it is based on only 8 SXW data points (1700+518 and
2112+059 do not have H$\beta$-shift measurements in BG92 due
to their weak [O~{\sc iii}] emission), it does appear to be
reasonably robust. If we add a hypothetical SXW data point 
with an H$\beta$ shift of 0, we still find a significant difference 
between the H$\beta$-shift distributions of SXW and non-SXW. 
In this case, the Kuiper test probability is 0.0132. 

SXW~QSOs tend to lie toward the weak-[O~{\sc iii}] end of 
`eigenvector~1' as defined by BG92. Eigenvector~1 
describes a set of optical emission-line properties 
that vary together in a highly coordinated manner 
(e.g., [O~{\sc iii}] strength, Fe~{\sc ii} strength, H$\beta$ FWHM, 
and H$\beta$ asymmetry), and it emerged from the Principal Component 
Analysis (PCA) performed by BG92. Note, however, that eigenvector~1 
was derived from a PCA that included \aox\ (see Table~4 of BG92). In 
order to be sure that the weak dependence of eigenvector~1 upon \aox\  
is not affecting our result, we have performed our own PCA with \aox\ 
excluded. We find that our SXW~QSOs still tend to lie at the 
weak-[O~{\sc iii}] end of this `\aox-free' eigenvector~1 (see Table~3).

\subsection{Radio Structures}

The radio structures of SXW and non-SXW are more difficult to 
compare in a rigorous statistical manner due to the difficulty
involved in quantifying radio structure. Our one radio-loud 
SXW~QSO, 1004+130, is lobe dominated, suggesting that we are
viewing the active nucleus in a relatively edge-on manner
(see Wills, Brandt \& Laor 1999 for detailed discussion). 
There are no obvious radio structure differences between the 
radio-quiet members of SXW and non-SXW, although we note that 
better radio imaging data are needed to properly examine this 
matter. The only radio-quiet SXW~QSO with radio structure of 
particular note is 1700+518. It shows a double source morphology 
(e.g., see \S5.1.3 of Miller et~al. 1993 and \S3.1 of 
Kukula et~al. 1998). 

2251+113 is radio-loud and fairly soft X-ray weak (see \S6), 
and it also is lobe dominated.


\section{Ultraviolet Absorption-Line Properties of Soft X-ray Weak QSOs}

If photoelectric X-ray absorption is the primary cause of soft X-ray 
weakness for QSOs, then the X-ray absorbing material might also be 
detected at other wavelengths. The obvious place to look is in the
ultraviolet where cosmically abundant elements have strong
bound-bound transitions (e.g., Mathur et~al. 1994). Alternatively, 
X-ray and ultraviolet absorption might have a high frequency of joint 
occurrence even if the two types of absorption do not arise in the same 
gas (e.g., Crenshaw et~al. 1998 and references therein). 

We have therefore collected the available \hst\ and \iue\ 
spectra for the BG92 QSOs to examine if SXW~QSOs preferentially show 
strong ultraviolet absorption. We have focused on absorption in the 
C~{\sc iv}~$\lambda 1549$ line (hereafter C~{\sc iv}) because 
the most data are available for this region. There is a 
wide range in the quality of the spectra, and spectra with particularly 
poor signal to noise were rejected (see below). Preference was usually given 
to \hst\ data. We found 46 objects with adequate \hst\ coverage of at least 
the C~{\sc iv} region and 9 additional objects with adequate \iue\ 
coverage. When several datasets were available we co-added them, applying 
any needed shifts in wavelength scale and weighting appropriately. The 
rest wavelength scale was determined from a redshift based on the best 
[O~{\sc iii}]~$\lambda 5007$ data where possible.

We have searched all the resulting spectra for BALs or other associated 
C~{\sc iv} absorption features, and in Table~1 we give our measured 
rest-frame EWs. All measurements were performed by one author with 
substantial experience in \hst\ and \iue\ data analysis (BJW), and 
particular effort was made to ensure consistency and uniformity.
The measurements were made independently of knowledge of \aox.  
We excluded narrow Galactic interstellar absorption lines 
based on line identifications over the available spectral range. 
We were able to confirm some weak C~{\sc iv} absorption features 
from the doublet wavelength separations or the presence of 
corresponding absorption features (e.g., N~{\sc v}~$\lambda 1240$, 
Ly$\alpha$, O~{\sc vi}~$\lambda 1034$) in the same redshift system. 
Generally, only measurements with signal to noise greater than 4 
were retained. When no absorption was detected we enter 0 for the EW.
While our EWs refer to measurable features, they are probably also a
reasonable measure of the total EW of C~{\sc iv} associated absorption 
for a given object. However, we do not give detailed uncertainties. 
This is because it is difficult to give meaningful limits to any 
unmeasured associated absorption. Such limits are a complicated 
function of many parameters: 
noise, 
the unknown widths of possible absorption features, 
the effective continuum placement (especially in the presence of 
broad emission line wings), and 
the actual level of the local continuum (which may be low in the 
true continuum to high near the center of a broad emission line). 
The archival data are readily accessible to the interested reader, 
and some examples of spectra are given in \S5 and \S6.

In Figure~4 we show a plot of \aox\ versus C~{\sc iv} absorption 
EW. This plot reveals a striking connection between soft X-ray 
weakness and strong C~{\sc iv} absorption. Note that the ordinate 
of this plot spans a range of $\simgt 800$ in C~{\sc iv} absorption 
EW. We have labeled our SXW~QSOs and other QSOs of particular 
interest, and we give object-specific notes on these QSOs in \S5
and \S6. At least eight of our ten SXW~QSOs show C~{\sc iv} 
absorption with EW~$>1$~\AA, and most have EW~$>5$~\AA. 
We have performed a Spearman rank-order correlation analysis using 
the data shown in Figure~4 (including appropriate censoring; 
Isobe, Feigelson \& Nelson 1986) and find the correlation to be 
significant with $>99$\% confidence. We have searched for other
correlations that could be inducing the correlation of Figure~4 
as a secondary effect, and we find none. For example, we do not
find any correlation between H$\beta$ FWHM and C~{\sc iv} 
absorption EW. This correlation is worth checking because 
$\alpha_{\rm x}$ depends upon H$\beta$ FWHM (see \S2); this
relationship could have induced a weak dependence of \aox\ 
on H$\beta$ FWHM.

In Figure~4 we have also marked bona-fide BAL~QSOs with 
asterisks.\footnote{To be spoken of as a 
bona-fide BAL~QSO, we require a QSO to satisfy the non-zero
balnicity requirement outlined in \S3.1 of 
Weymann et~al. (1991). To our knowledge, the only BG92 QSOs 
that clearly satisfy this requirement are 
0043+039,
1001+054,  
1700+518, 
2112+059 and probably  
1004+130 (see \S5).} 
These 5 BAL~QSOs have the largest C~{\sc iv} absorption EW values in
our SXW~QSO sample, and inspection of Table~2 reveals that they are
also our 5 most optically luminous SXW~QSOs. In light of this result, 
we have used a Spearman analysis to search for a correlation between
$M_{\rm V}$ and C~{\sc iv} absorption EW for our 10 SXW~QSOs, and we 
find a correlation that is significant with $>99$\% confidence.
However, with only 8 data points and 2 upper limits we cannot
rigorously demonstrate a causative relation between $M_{\rm V}$ 
and C~{\sc iv} absorption EW for our SXW~QSOs.

Finally, we have used a Spearman analysis to search for a correlation 
between the C~{\sc iv} absorption EW and the broad-band optical continuum 
polarization (again from Berriman et~al. 1990) for our 55 BG92 QSOs
with C~{\sc iv} data. The analysis indicates a correlation to be 
present at the $>96$\% confidence level, but a plot of these two 
quantities shows that any correlation has a large scatter. 


\section{Notes on Individual Soft X-ray Weak QSOs}

In this section we give relevant comments on some of our SXW~QSOs.
In Figure~5 we show spectra of the C~{\sc iv} region for our 
SXW~QSOs, and the reader should refer to this figure while reading 
this section. 0043+039 and 1700+518 are well-known BAL~QSOs that have 
been intensively studied, so we shall not give comments on them here.


{\bf 1001+054:}    
This QSO was first noted to be soft X-ray weak by Laor et~al. (1997),
and it has a narrow-line type~1 spectrum. \hst\ spectra 
reveal broad C~{\sc iv} and Ly$\alpha$ absorption, and we measure a
small, but non-zero, balnicity index (see Weymann et~al. 1991) 
of $\approx 420$~km~s$^{-1}$. A detailed study of the \hst\ data will 
be presented in Wills et~al., in preparation. 


{\bf 1004+130 (PKS~1004+130, 4C~13.41):} 
This is the only radio-loud SXW~QSO in our sample ($\log R=2.36$ and the
radio spectrum is steep). 
Its polarization percentage rises with frequency, reaching
$\approx 2$\% by 4000~\AA\ (Antonucci et~al. 1996 and references therein). 
It was first noted to be soft X-ray weak by Elvis \& Fabbiano (1984),
who studied it using data from \einstein\ and \iue. In the low-resolution 
\iue\ spectra Kinney et~al. (1991) recognized unresolved 
N~{\sc v} and C~{\sc iv} absorption near the emission-line redshift.
A high-resolution \hst\ GHRS spectrum was obtained by Bowen et~al. (1997) 
to search for intervening absorption by the low-redshift dwarf spheroidal 
galaxy Leo~I. Serendipitously, this 
spectrum showed the associated N~{\sc v} absorption 
doublet with at least two velocity components (Figure~3 of Bowen et~al. 1997).  
In this same spectrum, which we retrieved from the \hst\ archives, we 
discovered the corresponding high-ionization absorption lines of 
the O~{\sc vi} doublet (see Figure~6). These appear to be optically thick and
suggest the presence of partial covering. In addition, our detailed analyses
of the \iue\ and \hst\ data suggest much broader unresolved troughs of
C~{\sc iv}, Si~{\sc iv} and O~{\sc vi}, extending to outflow velocities 
of $\approx 10\,000$~km~s$^{-1}$. We estimate a balnicity index of
$\approx 850$~km~s$^{-1}$ and believe that 1004+130 is probably a 
low-redshift, radio-loud BAL~QSO. There is probable Ly$\alpha$ absorption 
and strong N~{\sc v} absorption too. This absorption affects the 
Ly$\alpha$ emission line and accounts for its weakness in this QSO. A 
detailed analysis of the ultraviolet data is presented by 
Wills, Brandt \& Laor (1999). 

  
{\bf 1011--040:}  
The archival \iue\ data for this SXW~QSO do not reveal any obvious 
associated absorption, but the spectral quality is not high.
It seems unlikely but not impossible that the blue wing of the 
C~{\sc iv} emission line could be stronger than observed, implying 
a larger absorption EW. In this case the observed emission 
line could appear redshifted. However, the peaks of 
the Ly$\alpha$ and C~{\sc iv} emission lines give a consistent 
redshift, and the symmetry of the stronger Ly$\alpha$ line 
does not hint at significant absorption.


{\bf 1126--041 (Mrk~1298):}   
Wang et~al. (1999a) have recently studied the ultraviolet and X-ray
absorption in the spectrum of this QSO. \hst\ spectra of this object 
would allow the strong ultraviolet absorption to be studied in 
significantly more detail. 


{\bf 1411+442:}    
1411+442 was noted to be soft X-ray weak by Laor et~al. (1997), 
and its ultraviolet absorption and other properties are discussed in 
detail by Malkan, Green \& Hutchings (1987) and Wang et~al. (1999b). 
While its ultraviolet absorption is prominent, it is 
not a bona-fide BAL~QSO. 
Brinkmann et~al. (1999) have presented the \asca\ data for this
QSO, and our independent analysis of these data is in general 
agreement with theirs. The data suggest absorption by an intrinsic 
column density of $\approx$~(1--2)$\times 10^{23}$~cm$^{-2}$, and 
there is also evidence for a scattered X-ray component at low energies. 
The \asca\ spectrum has limited photon statistics, and the X-ray 
column density is subject to significant uncertainty because
the underlying continuum shape and the ionization state of the 
X-ray absorber cannot be tightly constrained. 


{\bf 1535+547 (Mrk~486):} 
This object is our weakest detected SXW~QSO, and it is also 
the most highly polarized BQS QSO. Its polarization (2--8\%) and 
associated C~{\sc iv} absorption have recently been studied by 
Smith et~al. (1997), and they argue for a complex 
broad emission-line region that contains dust.
1535+547 has a narrow-line type~1 spectrum.


{\bf 2112+059:}    
The \hst\ spectrum for this SXW~QSO reveals broad ultraviolet absorption 
by C~{\sc iv} and Ly$\alpha$ (Jannuzi et~al. 1998). The broad absorption 
tends to be shallow, but there are narrower absorption components as well. 
We measure a balnicity index of $\approx 2980$~km~s$^{-1}$. 
2112+059 also has an intervening damped Ly$\alpha$ absorption system at 
$z=0.2039$ (e.g., Lanzetta, Wolfe \& Turnshek 1995), but this system does 
not materially affect our analysis of the associated ultraviolet 
absorption. We do not expect this system to cause substantial neutral 
or ionized X-ray absorption; its neutral hydrogen column density is
$\approx 2.5\times 10^{20}$~cm$^{-2}$, and we expect any ionized
column density to be $\simlt 5\times 10^{19}$~cm$^{-2}$. 


{\bf 2214+139 (Mrk~304):}   
The archival \iue\ spectrum for this SXW~QSO does not reveal any 
obvious associated absorption, but the spectral quality is not 
high (also see Clavel \& Joly 1984). It is not impossible that 
the blue wing of the C~{\sc iv} emission line could be stronger 
than it appears, hiding significant absorption.  However, the 
stronger, symmetric Ly$\alpha$\ emission line shows no hint of 
corresponding blueshifted absorption.
Rachen et~al. (1996) noted that the flatness of the \rosat\ PSPC 
spectrum might be due to photoelectric absorption, but detailed 
spectral analysis of these data is not possible due to the limited 
photon statistics.


\section{Notes on Other Relevant QSOs}

Here we give comments on other objects that lie in interesting 
locations in Figure~4. The reader should refer to this figure 
while reading this section. 


{\bf 0844+349:}  
Yuan et~al. (1998) found 0844+349 to be a SXW~QSO during a pointed 
\rosat\ observation but not during the RASS. During the pointed
observation they found $\alpha^{\prime}_{\rm ox}=-2.05$ (calculated 
between 2500~\AA\ and 2~keV). For these data we calculate 
$\alpha_{\rm ox}=-1.83$ using the methods in \S2, and thus
0844+349 does not quite satisfy our criterion for a SXW~QSO. 
This QSO was also not soft X-ray weak during an \einstein\ IPC 
observation (Kriss 1988). The current data suggest that 0844+349 is 
not usually a SXW~QSO, and it appears to have been caught in an 
unusually low flux state during the pointed \rosat\ observation.
It has shown soft X-ray variability by a factor of $\approx 6$,
and this may be due to time-variable X-ray absorption. The poor
statistics of the pointed \rosat\ spectrum make it difficult
to examine this issue in detail, especially if significant X-ray
scattering is present.
Corbin \& Boroson (1996) comment that this QSO shows `associated absorption'
in an \hst\ spectrum, and our analysis of the \hst\ data they used reveals 
narrow associated Ly$\alpha$ absorption. There are no \hst\ data for the 
C~{\sc iv} region, but co-added \iue\ data suggest C~{\sc iv} absorption 
consistent in velocity with the Ly$\alpha$ absorption. 
%


{\bf 1114+445:}  
This QSO shows moderate strength X-ray and ultraviolet absorption by 
ionized gas (George et~al. 1997; Mathur, Wilkes \& Elvis 1998). Its
intermediate position relative to the low-absorption QSOs and  
SXW~QSOs in Figure~4 is notable. 1114+445 is one of the more 
highly polarized BQS QSOs (2.3\%; Berriman et~al. 1990). 


{\bf 1259+593:} 
As noted in \S2, this object has only a poor limit on its X-ray flux, 
and we cannot at present determine if it satisfies our 
$\alpha_{\rm ox}\leq-2$ criterion to be a SXW~QSO. Bahcall et~al. (1993) 
presented an \hst\ spectrum for this object, and there is no 
obvious intrinsic absorption (also see Lu \& Zuo 1994). 
We do not expect significant X-ray absorption from high-velocity cloud 
Complex~$C$ since its neutral hydrogen column density in this direction
is $\approx 1.5\times 10^{20}$~cm$^{-2}$ (see Savage et~al. 1993 and 
references therein). Similarly, UGC~8040 and UGC~8046 should not
cause significant X-ray absorption given the ultraviolet constraints 
on neutral hydrogen absorption lines and the possible ionization 
levels in the interstellar media of these galaxies (see 
Bowen et~al. 1996 and references therein). 


{\bf 1309+355:} 
This is a flat-spectrum radio-intermediate QSO ($\log R=1.26$), and 
Miller et~al. (1993) and Falcke et~al. (1996) argue 
that it is a boosted radio-quiet QSO. The radio 
properties may indicate that the central engine is viewed in a relatively 
pole-on manner. In light of this, the strong ultraviolet absorption lines 
seen in the \hst\ spectrum (C~{\sc iv}, N~{\sc v}, O~{\sc vi} and  Ly$\alpha$) 
are notable. This object has $\alpha_{\rm ox}=-1.71$; it is moderately 
X-ray weak, especially for a radio-loud QSO. 


{\bf 1351+593:} 
This QSO is moderately weak in soft X-rays (also see Elvis 1992), and it shows 
associated ultraviolet absorption in \hst, \hut\ and \iue\ spectra
(Brosch \& Gondhalekar 1984; Granados et~al. 1993; Zheng et~al. 1999). 
The absorption runs to fairly high velocities ($\approx 3000$~km~s$^{-1}$),
but 1351+593 probably does not qualify as a bona-fide BAL~QSO.


{\bf 1402+261:} 
This QSO shows moderate strength ultraviolet absorption by Ly$\alpha$ 
and C~{\sc iv}. The available \rosat\ data are consistent
with no intrinsic X-ray absorption by either neutral or ionized gas
(Ulrich-Demoulin \& Molendi 1996; Laor et~al. 1997). 



{\bf 1404+226:} 
1404+226 has one of the most extreme narrow-line type~1 spectra
for a QSO, and its X-ray spectrum shows a strong soft X-ray
excess as well as additional poorly understood complexity 
(e.g., Ulrich-Demoulin \& Molendi 1996; Leighly et~al. 1997;
Ulrich et~al. 1999). The additional complexity may be associated 
with X-ray absorption edges or lines. \hst\ spectra for 1404+226 
reveal moderate strength ultraviolet absorption
(Ulrich et~al. 1999).  


{\bf 1425+267:} 
Laor et~al. (1997) pointed out that 1425+267 was moderately soft
X-ray weak compared to other radio-loud QSOs, and it has a clear
double-lobed radio structure. Complex residuals 
below 1~keV in archival \asca\ spectra suggest the 
presence of intrinsic X-ray absorption, although the 
absorption cannot be probed in detail due to limited photon
statistics. The \hst\ spectrum for 1425+267 shows 
moderate-strength absorption by Ly$\alpha$ and C~{\sc iv}. 


{\bf 1552+085:} 
Turnshek et~al. (1997) suggest that this QSO has a C~{\sc iv} BAL
in its \iue\ spectrum, but the data are noisy and better spectra 
are needed to reliably study ultraviolet absorption. For this 
reason, we do not quote a C~{\sc iv} absorption EW in Table~1, and 
this object does not appear in Figure~4. We find an \aox\ value 
of $-1.77$, which is fairly small but still somewhat larger 
than is typically seen for BAL~QSOs. 1552+085 is one of the 
more highly polarized BQS QSOs (1.9\%; Berriman et~al. 1990). 


{\bf 1704+608 (3C351):}  
This is a lobe-dominated, radio-loud QSO ($\log R=2.81$) with 
moderate strength X-ray and ultraviolet absorption by ionized 
gas (e.g., Mathur et~al. 1994; Nicastro et~al. 1999). Like 
1114+445 and 1425+267, it occupies an intermediate position 
between low-absorption QSOs and our SXW~QSOs (also see 
Fiore et~al. 1994). 1704+608 does not have notable polarization. 


{\bf 2251+113:} 
This steep-spectrum radio-loud QSO ($\log R=2.56$) is moderately weak 
in the soft X-ray band, and associated ultraviolet absorption by 
Ly$\alpha$, N~{\sc v}, C~{\sc iv} and Si~{\sc iv} has been
detected by \hst\ (Bahcall et~al. 1993). We have analyzed
an archival \asca\ spectrum of this QSO, and these data suggest
absorption by an intrinsic column density of 
$\simgt 3\times 10^{21}$~cm$^{-2}$. However, as for 
1411+442 and 1425+267, the \asca\ spectrum has limited photon 
statistics and hence the X-ray column density is subject to 
significant uncertainty.


\section{\boldmath$\alpha_{\rm ox}$ Correlations after Removal of the Soft X-ray Weak QSOs}

We have used our improved \aox\ values for the BG92 QSOs to examine
correlations between \aox\ and other observables in the most general
manner possible. In these correlation analyses, we have excluded our 
10 SXW~QSOs from consideration. This is important since here we are 
interested in physical correlations that are not affected by objects
where \aox\ has been altered by strong X-ray absorption (see \S8 for 
the evidence that soft X-ray weakness is primarily due to absorption). 
We have performed Spearman rank-order correlations between our \aox\ 
values and the quantities tabulated by BG92, and our results are given 
in Table~4. 
In agreement with Green (1998), we find significant 
($P_{\rm Spearman}<0.04$) correlations between \aox\ 
and the EWs of H$\beta$, [O~{\sc iii}], He~{\sc ii}
and Fe~{\sc ii}. 
In addition, we find the \aox\ correlations against 
He~{\sc ii}/H$\beta$, 
Fe~{\sc ii}/H$\beta$,
H$\beta$ FWHM,  
H$\beta$ asymmetry, 
$\alpha_{\rm opt}$, 
eigenvector~1,  
eigenvector~2 (see \S3.2 of BG92 for discussion), and
the `\aox-free' eigenvector~1 
to be significant. 
The absence of a significant correlation with $R$ is notable. Earlier 
studies have found a correlation between \aox\ and $R$ (e.g., 
Zamorani et~al. 1981, hereafter Z81), and this apparent discrepancy 
is discussed below in \S8.2. 
Our correlations supersede and extend those of BG92 since
(1) our \aox\ values are more complete and constraining than those used 
by BG92 and (2) we have removed objects with evidence for strong  
X-ray absorption (see above).


\section{Discussion and Conclusions}

\subsection{The Origin and Demography of Soft X-ray Weak QSOs}

We have systematically investigated the nature of the 
SXW~QSOs in the $z<0.5$ BQS. On the whole, the data support 
the idea that X-ray absorption is the primary 
cause of soft X-ray weakness in QSOs. We detect remarkably strong 
(EW~$>4.5$~\AA) C~{\sc iv} absorption in 8 of our 10 SXW~QSOs, and 
ultraviolet and X-ray absorption have a high probability of joint
occurrence. For comparison, only 1 of 45 non-SXW QSOs with 
C~{\sc iv} coverage shows absorption with EW~$>4.5$~\AA. 
The two SXW~QSOs without clear ultraviolet 
absorption, 1011--040 and 2214+139, both have \iue\ spectra of only 
limited quality, and \hst\ spectra are needed to further examine the 
possibility of ultraviolet absorption.
\asca\ data are available for only three of our SXW~QSOs: 0043+039, 
1411+442 and 1700+518. The SXW~QSOs 0043+039 and 1700+518 were not 
detected (consistent with the presence of heavy absorption; see 
Gallagher et~al. 1999), and 1411+442 shows evidence for substantial 
absorption in its X-ray spectrum (see \S5). 

Furthermore, we have identified a general correlation between 
\aox\ and C~{\sc iv} absorption EW that appears to be due to a 
continuum of absorption properties (see Figure~4). Unabsorbed QSOs 
and SXW~QSOs lie towards opposite ends of this correlation, while 
QSOs with X-ray warm absorbers lie near the middle of the 
correlation (see the extensive object notes in \S5 and \S6). 
BAL~QSOs, which comprise a subclass of SXW~QSOs, appear to 
lie at the absorbed extreme of the correlation; the 
bona-fide BAL~QSOs in our SXW~QSO sample have ultraviolet
absorption that is the strongest we observe, and they often
have only upper limits on their X-ray fluxes. 
The observed correlation is generally consistent with models that 
postulate a connection between orientation and absorption strength. 
For example, as one increases the inclination angle one might move 
through the sequence: unabsorbed QSO, X-ray warm absorber QSO, 
non-BAL SXW~QSO, BAL SXW~QSO, and perhaps type~2 QSO. 
However, our results in \S8.2 regarding [O~{\sc iii}] luminosity
suggest that this picture cannot be complete and that an intrinsic 
property must also play an important role. In addition, we note
that there is evidence suggesting the central engine in the
moderately soft X-ray weak QSO 1309+355 is viewed in a relatively 
pole-on manner (see \S6). Finally, the potential relation between 
$M_{\rm V}$ and C~{\sc iv} absorption EW for our SXW~QSOs (see \S4)
would be difficult to understand in the context of a pure orientation
model. 

Although the correlation in Figure~4 strongly suggests that soft 
X-ray weakness is due to absorption, and that the X-ray and ultraviolet 
absorbers are related, this correlation does not imply that these 
absorbers are identical. In fact, a uniform screen which completely 
covers both the X-ray and ultraviolet emission sources is 
not expected to produce the shape of the correlation seen in Figure~4. 
This follows since the ultraviolet absorption is by a resonance line, 
where the absorption EW grows with the absorbing column $N_{\rm H}$ 
following a `curve of growth' (e.g., Chapter~14 of Gray 1992), while 
the X-ray absorption is by bound-free edges, and thus grows 
like $e^{-\tau_{\rm bf}}$ where $\tau_{\rm bf}\propto N_{\rm H}$. 
One expects \aox\ to be practically independent
of the C~{\sc iv} EW up to an $N_{\rm H}$ which gives $\tau_{\rm bf}\sim 1$,
and above this one expects a very rapid drop in \aox\ associated
with a very slow increase in the C~{\sc iv} EW. Instead, Figure~4 
indicates a gradual increase in the C~{\sc iv} EW is associated with 
a gradual decrease in \aox. This may be explained
if both the X-ray and ultraviolet absorption are optically 
thick, but with an absorption covering factor which is less than unity 
(possibly due to scattering). In this case both \aox\ and the 
C~{\sc iv} EW just provide an indication of the X-ray and ultraviolet 
absorber covering factors, which may vary together.

Other conceivable causes of soft X-ray weakness include 
(1) an unusual underlying spectral energy distribution and 
(2) extreme X-ray or optical variability.
However, we consider it unlikely that either of these 
is the primary cause of soft X-ray weakness in optically
selected QSOs. 
Regarding possibility~1, SXW~QSOs might be postulated to be 
intrinsically weak emitters of soft X-rays. Our optical line studies 
described in \S3 allow us to address this since line EWs depend upon 
the shape of the ionizing spectral energy distribution. We do not, for 
example, find evidence for a statistically significant difference in the 
He~{\sc ii}~EW distributions for SXW~QSOs and non-SXW~QSOs. The 
creation of the He~{\sc ii} line is driven by 54--150~eV continuum 
photons, and it would thus be difficult to produce a normal strength 
line if the underlying spectral energy distribution were anomalously 
weak in this energy range (see Korista, Ferland \& Baldwin 1997 
and references therein). 

Regarding possibility~2 of the previous paragraph, extreme X-ray 
variability (with variability amplitude $\simgt 10$) has been 
seen for some BQS~QSOs, while such extreme optical variability 
appears much rarer (e.g., Giveon et~al. 1999). To investigate 
potential \aox\ variability among our 10 SXW~QSO, we have 
compared our \aox\ measurements and upper limits with the 3 \aox\ 
measurements and 3 \aox\ upper limits of 
Tananbaum et~al. (1986; see Table~2). While the data are limited, 
we do not find evidence for outstanding \aox\ variability.  

Our results imply that selection by soft X-ray weakness 
is an effective ($\simgt 80$\% successful)
way to find low-redshift QSOs with strong ultraviolet 
absorption. This is important from a practical point of view 
because the optical and X-ray flux densities needed to establish soft 
X-ray weakness can be obtained in an inexpensive manner from publicly 
available optical images (e.g., the Palomar Schmidt and UK Schmidt 
sky surveys) and X-ray data (e.g., the RASS and 
\rosat\ pointed observations). When data from the RASS are used, 
we expect this method to be effective down to about $B=17$; for
(unabsorbed) QSOs with lower optical fluxes, the expected X-ray flux 
becomes comparable to or less than the RASS sensitivity limit. This 
selection method appears to be significantly more effective than 
others that have been developed to find QSOs with strong ultraviolet 
absorption (e.g., Turnshek et~al. 1997). 

We have found the percentage of SXW~QSOs in the optically selected 
QSO population (specifically, the $z<0.5$ BQS) to be $\approx 11$\%. 
This percentage is in general agreement with previous rough estimates 
but is more statistically reliable since it is derived from a larger 
and better-defined sample.
We find 4--5 bona-fide BAL~QSOs (0043+039, 1001+054, 1700+518, 2112+059 
and probably 1004+130) in our sample, and if all BAL~QSOs are soft 
X-ray weak then our methods should have identified all the BAL~QSOs in 
the $z<0.5$ BQS. It is of interest to examine if the BAL incidence 
we find is consistent with that found in other QSO 
surveys (see \S11.1 of Krolik 1999). 
Weymann (1997) gives a BAL~QSO incidence of about $11$\% for 
the $z=$~1.4--3.0 QSOs from the Large Bright Quasar Survey 
(LBQS; see Hewett, Foltz \& Chaffee 1995),
and we would have expected to detect $\approx 9.6$ BAL~QSOs given this
percentage.\footnote{We believe the `true' 11\% incidence rather than the
`raw' 8\% incidence is the appropriate number to use for comparison
in this case (see Weymann 1997). At the wavelengths at which they were 
selected, the $z<0.5$ BQS QSOs do not have their fluxes 
diminished by BAL troughs.} The number of BALQSOs we detect, while 
perhaps somewhat smaller, does not show any strong inconsistency 
with the expectation from the LBQS. A Fisher exact probability test 
for a $2\times 2$ contingency table gives a probability of $\approx 5$\% 
when we compare the BQS and LBQS samples. 
Our results are in general agreement with the idea that the 
incidence of BALs in modest-redshift QSOs is 
about the same as in higher-redshift QSOs. 

As mentioned in \S4, the 5 bona-fide BAL~QSOs in our SXW~QSO sample
are our 5 most optically luminous SXW~QSOs. They also have the
largest C~{\sc iv} absorption EWs, and among our SXW~QSOs we find
suggestive evidence for a correlation between $M_{\rm V}$ and 
C~{\sc iv} absorption EW. If the outflows from our SXW~QSOs 
are optically thick, the C~{\sc iv} absorption EW is expected
to be proportional to just the outflow velocity spread and the outflow
covering factor, and the outflow velocity spread is plausibly 
related to the ultraviolet luminosity (e.g., see equation~6 of 
Scoville \& Norman 1995). Thus a relation between $M_{\rm V}$ 
and C~{\sc iv} absorption EW seems physically plausible, but 
further work is clearly needed to examine the reality of any relation. 

\subsection{Interpretation of the Continuum and Optical Emission-Line Analyses}

The continuum and optical emission-line analyses discussed in \S3 show
that SXW~QSOs do differ in some respects from non-SXW~QSOs. 
The low [O~{\sc iii}] luminosities and EWs of SXW~QSOs are striking, 
and this phenomenon has been previously noted for a subset of 
SXW~QSOs, the low-ionization BAL~QSOs  
(e.g., Boroson \& Meyers 1992; Turnshek et~al. 1997). 
If [O~{\sc iii}] is an isotropic property, this result suggests 
that soft X-ray weakness is not merely caused by an orientation effect 
but arises at least in part as a result of an intrinsic 
property.\footnote{Here we are following the line of reasoning 
discussed in \S4.1 of BG92 and \S3.2 of Boroson \& Meyers (1992).
While we recognize that [O~{\sc iii}] may have some anisotropy
(e.g., see Hes, Barthel \& Fobsury 1996; di~Serego Alighieri et~al. 1997),
the best available evidence suggests that this anisotropy is not strong 
enough to remove the need for an intrinsic 
property (Kuraszkiewicz et~al. 1999a).} A clue to the nature of this
property may be found by noting that SXW~QSOs also lie toward
the weak-[O~{\sc iii}] `negative' end of BG92 
eigenvector~1 (see \S3).\footnote{Our only SXW~QSO
with a `positive' value of eigenvector~1 is the radio-loud QSO 1004+130,
and even this object has an unusually small value of eigenvector~1
for a radio-loud QSO. A somewhat similar connection between BAL~QSOs 
and strong optical Fe~{\sc ii} emitters has been discussed by 
Lawrence et~al. (1997), although the discussion there was based on 
qualitative argumentation with a fairly small sample of objects.} 
Many of the unabsorbed Seyferts and QSOs with similar values of 
eigenvector~1 have recently been found to have extreme X-ray spectral 
and variability properties, and it has been suggested that these 
`ultrasoft Narrow-Line Seyfert~1' objects are accreting at 
relatively high fractions of the Eddington 
rate ($\dot M/\dot M_{\rm Edd}$; e.g., BG92; 
Boller, Brandt \& Fink 1996; Laor et~al. 1997a). 
If an $\dot M/\dot M_{\rm Edd}$ interpretation of eigenvector~1 is 
correct, the similar eigenvector~1 values of SXW~QSOs would
suggest that they also have relatively high $\dot M/\dot M_{\rm Edd}$.
SXW~QSOs and ultrasoft Narrow-Line Seyfert~1s would be related objects that
differ primarily in the amount of absorption that happens to lie
along the line of sight. Accretion at high $\dot M/\dot M_{\rm Edd}$,
where radiation trapping effects are important, is expected to drive 
substantial mass outflow, and gas in such an outflow could produce 
the ultraviolet and X-ray absorption observed in SXW~QSOs (see \S5 of 
Blandford \& Begelman 1999). This gas, perhaps after having cooled, 
may also have been detected in ultrasoft Narrow-Line Seyfert~1 
objects via its emission lines; these objects 
appear to have especially large amounts of dense gas in their
nuclei (e.g., Kuraszkiewicz et~al. 1999b; Wills et~al. 1999). 
The significant differences in the H$\beta$ shifts and shapes of 
SXW~QSOs and non-SXW~QSOs suggest that either
(1) the dynamics of their Broad Line Regions (BLRs) are systematically different,
(2) particular parts of their BLRs are preferentially viewed, or
(3) broad H$\beta$ absorption near the line peak is modifying 
the profiles of SXW~QSOs (e.g., Anderson 1974). 
While such effects might plausibly be related to the outflows discussed above, 
we are at present unable to find a compelling interpretation for the 
shift and shape differences (this is largely due to the general lack of 
understanding of the structure and kinematics of the BLR). 

The absence of obvious optical continuum reddening, while initially 
somewhat surprising in light of the ultraviolet and X-ray absorption 
discussed above, can be understood if the dust in the absorbing gas 
has been destroyed by sputtering or sublimation. From an empirical
point of view, we also note that high-ionization BAL~QSOs show only
weak reddening despite their strong absorption (e.g., Weymann et~al. 1991). 
The generally low optical continuum polarization of our SXW~QSOs may 
also be understood by analogy with BAL~QSOs. While scattering 
by electrons or dust, together with dust absorption of the central continuum,
has been used to explain the high polarization of some BALQSOs, about 
half of BAL~QSOs have polarization percentages $\simlt 1$\% 
(e.g., Schmidt \& Hines 1999).

An apparently surprising result from \S7 is that \aox\ does not appear 
to correlate with $R$. Earlier studies, in contrast, have found that for 
a given optical luminosity, the average X-ray emission of radio-loud 
QSOs is $\sim 3$ times higher than that of radio-quiet QSOs
(e.g., Z81). 
This apparent discrepancy is due to the following three effects: 
(1) Z81 define radio-loud QSOs as those having $\log R>1.88$ rather than $\log R>1$,
(2) Z81 do not detect QSOs with $\alpha_{\rm ox}<-1.75$ (see their Figure~4), and 
(3) the radio-loud QSOs in Z81 extend up to $\log R\sim 4.8$, while ours have
$\log R\simlt 3$.
Figure~5 of Z81 suggests a flattening of \aox\ with increasing $\log R$, and to
make a proper comparison with the BG92 radio-loud QSOs one needs to take 
subsamples from both samples which follow the same selection criteria. The
average \aox\ for the subsample of 16 QSOs from Table~1 of Z81 with $1\le\log R\le 3$
is $\langle \alpha_{\rm ox}\rangle=-1.48\pm 0.05$, 
which is about equal to the Z81 mean for the radio-quiet QSO population 
($-1.46\pm 0.06$) and our mean of the subsample of 14 
$\alpha_{\rm ox}>-1.75$ radio-loud QSOs from BG92 ($-1.48\pm 0.11$).
Thus, there is no discrepancy between the Z81 results and our 
results. The difference in \aox\ between radio-quiet and radio-loud samples 
appears to be driven primarily by the radio-loudest ($\log R>3$) QSOs, 
which are not present in our sample. 

\subsection{Future Studies}

Systematic hard X-ray spectroscopy of the SXW~QSOs discussed in 
this paper is the obvious next step in the study of these objects. 
Such spectroscopy should allow determination of the column densities, 
ionization parameters, and covering factors of the expected X-ray 
absorbers, and it would let one critically examine the continuum of 
absorption properties suggested by the \aox\ versus C~{\sc iv} 
absorption EW correlation. 
Furthermore, improved ultraviolet spectra are needed for several 
of our SXW~QSOs. Absorption in species in addition to C~{\sc iv} 
can constrain physical conditions in the ultraviolet absorber. 
Objects such as 1011--040 and 2214+139 have only
weak limits on C~{\sc iv} absorption and might well show interesting 
ultraviolet absorption in higher quality spectra. Studies of these two
objects will determine whether X-ray weakness in the BG92~QSOs is always 
associated with ultraviolet absorption, or whether it is possible in 
some cases to get X-ray absorption without noticeable ultraviolet 
absorption. Objects such as 1004+130 and 1126--041 also need better 
spectra to constrain the geometry and dynamics of their known 
ultraviolet absorption. 
We are working to obtain the required X-ray and ultraviolet data
for our sample.  

Finally, the soft X-ray weakness selection method described 
above may be profitably applied to larger QSO samples. 


\acknowledgments

We thank 
Th. Boller, 
M. Elvis, 
E. Feigelson, 
J. Nousek, and
D. Schneider
for helpful discussions. 
We thank T. Boroson for providing data from BG92. 
We gratefully acknowledge the support of 
NASA LTSA grant NAG5-8107 and the Alfred P. Sloan Foundation (WNB),
the fund for the promotion of research at the Technion (AL), and
NASA LTSA grant NAG5-3431 (BJW). 



\begin{deluxetable}{llllll}
\tablenum{1}
\tablewidth{0pt}
\tablecaption {\aox\ and C~{\sc iv} absorption EW values for QSOs in the BG92 sample$^1$}
\small
\tablehead{
\colhead{PG}                   & 
\colhead{\aox/}                &     
\colhead{PG}                   & 
\colhead{\aox/}                &     
\colhead{PG}                   & 
\colhead{\aox/}\\        
\colhead{Name}                 & 
\colhead{C~{\sc iv} EW (\AA)}  &     
\colhead{Name}                 & 
\colhead{C~{\sc iv} EW (\AA)}  &     
\colhead{Name}                 & 
\colhead{C~{\sc iv} EW (\AA)}         
}
\startdata
$0003+158$ & $-1.38$/0.8              & $1115+407$ & $-1.45$/0.4       & $1425+267$ & $-1.63$/1.8              \nl
$0003+199$ & $-1.50$/0                & $1116+215$ & $-1.57$/0.1       & $1426+015$ & $-1.46$/0                \nl 
$0007+106$ & $-1.43$/$\cdots^{\rm a}$ & $1119+120$ & $-1.58$/$\cdots$  & $1427+480$ & $-1.52$/0.03             \nl  
$0026+129$ & $-1.50$/0                & $1121+422$ & $-1.59$/0         & $1435-067$ & $-1.63$/$\cdots$         \nl  
$0043+039$ & $<-2.00$/$15.2^{\rm b}$  & $1126-041$ & $-2.13$/5.4       & $1440+356$ & $-1.38$/0                \nl 
$0049+171$ & $-1.27$/$\cdots$         & $1149-110$ & $-1.42$/$\cdots$  & $1444+407$ & $-1.57$/$\cdots^{\rm a}$ \nl
$0050+124$ & $-1.56$/0.4              & $1151+117$ & $-1.46$/$\cdots$  & $1448+273$ & $-1.59$/$\cdots$         \nl
$0052+251$ & $-1.37$/0                & $1202+281$ & $-1.27$/0.4       & $1501+106$ & $-1.64$/$\cdots$         \nl 
$0157+001$ & $-1.60$/$\cdots$         & $1211+143$ & $-1.57$/0.5       & $1512+370$ & $-1.43$/0.2              \nl 
$0804+761$ & $-1.52$/$\cdots$         & $1216+069$ & $-1.44$/0         & $1519+226$ & $-1.51$/$\cdots$         \nl   
$0838+770$ & $-1.54$/$\cdots$         & $1226+023$ & $-1.47$/0         & $1534+580$ & $-1.38$/$\cdots$         \nl
$0844+349$ & $-1.54$/0.3              & $1229+204$ & $-1.49$/0         & $1535+547$ & $-2.45$/4.6              \nl
$0921+525$ & $-1.41$/$\cdots$         & $1244+026$ & $-1.60$/$\cdots$  & $1543+489$ & $-1.67$/0.4              \nl
$0923+129$ & $-1.41$/$\cdots$         & $1259+593$ & $<-1.75$/0        & $1545+210$ & $-1.38$/0                \nl
$0923+201$ & $-1.57$/$\cdots^{\rm a}$ & $1302-102$ & $-1.58$/0         & $1552+085$ & $-1.77$/$\cdots$         \nl
$0934+013$ & $-1.39$/$\cdots$         & $1307+085$ & $-1.52$/0         & $1612+261$ & $-1.41$/0                \nl
$0947+396$ & $-1.33$/0.20             & $1309+355$ & $-1.71$/2.8       & $1613+658$ & $-1.21$/0                \nl
$0953+414$ & $-1.50$/0.17             & $1310-108$ & $-1.52$/$\cdots$  & $1617+175$ & $-1.64$/$\cdots$         \nl
$1001+054$ & $-2.13$/12.6             & $1322+659$ & $-1.40$/0.2       & $1626+554$ & $-1.37$/0                \nl
$1004+130$ & $<-2.01$/11.5            & $1341+258$ & $-1.53$/$\cdots$  & $1700+518$ & $<-2.29$/84              \nl  
$1011-040$ & $-2.01$/$<1.2$           & $1351+236$ & $-1.52$/$\cdots$  & $1704+608$ & $-1.62$/2.4              \nl 
$1012+008$ & $-1.66$/$\cdots$         & $1351+640$ & $-1.78$/5.7       & $2112+059$ & $-2.11$/19               \nl   
$1022+519$ & $-1.34$/$\cdots$         & $1352+183$ & $-1.50$/0.5       & $2130+099$ & $-1.47$/0.8              \nl 
$1048+342$ & $-1.52$/$\cdots$         & $1354+213$ & $-1.39$/$\cdots$  & $2209+184$ & $-1.49$/$\cdots$         \nl 
$1048-090$ & $-1.41$/$\cdots$         & $1402+261$ & $-1.58$/1.2       & $2214+139$ & $-2.02$/$<1.2$           \nl   
$1049-006$ & $-1.56$/0.2              & $1404+226$ & $-1.55$/1.9       & $2233+134$ & $-1.66$/$\cdots$         \nl 
$1100+772$ & $-1.39$/0.6              & $1411+442$ & $-2.03$/8.5       & $2251+113$ & $-1.86$/0.8              \nl  
$1103-006$ & $-1.51$/0                & $1415+451$ & $-1.51$/0         & $2304+042$ & $-1.29$/$\cdots$         \nl 
$1114+445$ & $-1.62$/3.6              & $1416-129$ & $-1.56$/0         & $2308+098$ & $-1.35$/0.4              \nl
\enddata
\tablenotetext{1}{Absorption is considered associated if within 
$-12\,000$~km~s$^{-1}$ of the systemic redshift. We included absorption
line strengths beyond $-12\,000$~km~s$^{-1}$ only if the absorption
appeared to be associated with an absorption complex or BALs.
EW values are in the rest frame.}
\tablenotetext{a}{These QSOs have Ly$\alpha$ coverage but no C~{\sc iv} coverage. 
0007+106 and 0923+201 have Ly$\alpha$ absorption EWs of 0.5--1.0~\AA, while 1444+407 
has a Ly$\alpha$ absorption EW of $<0.5$~\AA.}
\tablenotetext{b}{This value was measured from Figure~10 of Turnshek et~al. (1994).}
\end{deluxetable}

\clearpage


\begin{deluxetable}{lcccc}
\tablenum{2}
\tablewidth{0pt}
\tablecaption {SXW~QSOs in the BG92 sample}
\small
\tablehead{
\colhead{PG Name}              & 
\colhead{$z$}                  & 
\colhead{$M_{\rm V}$}          & 
\colhead{Our \aox}             &       
\colhead{Tananbaum et~al. (1986) \aox$^{\rm a}$}       
}
\startdata
0043+039$^\star$ & 
0.384            &
$-26.16$         &
$<-2.00^{\rm b}$ &
$\cdots$         \nl
%
1001+054$^\star$     & 
0.161        &
$-24.07$     & 
$-2.13$      &       
$<-1.78$     \nl
1004+130$^{\rm \star c}$ & 
0.240              &
$-25.97$           & 
$<-2.01^{\rm d}$   &
$<-1.92$           \nl
1011--040    & 
0.058        &
$-22.70$     & 
$-2.01$      &
$-1.93$      \nl
%
1126--041    & 
0.060        &
$-23.00$     & 
$-2.13$      &       
$-1.90$      \nl
1411+442     & 
0.089        &
$-23.54$     & 
$-2.03$      &
$\cdots$     \nl
%
1535+547     & 
0.038        &
$-22.15$     & 
$-2.45$      &
$\cdots$     \nl
1700+518$^\star$ & 
0.292        &
$-26.44$     & 
$<-2.29$     &
$\cdots$     \nl
%
2112+059$^\star$     & 
0.466        &
$-27.26$     & 
$-2.11$      &       
$<-1.82$     \nl
2214+139     & 
0.067        &
$-23.39$     & 
$-2.02$      &
$-1.83$      \nl
%
%
%
\enddata
\tablenotetext{}{$^\star$The objects 0043+039, 1001+054, 1700+518, 
2112+059 and probably 1004+130 are BAL~QSOs.}
\tablenotetext{a}{These values were calculated using a power law defined 
by the rest-frame flux densities at 2500~\AA\ and 2~keV.}
\tablenotetext{b}{This \aox\ value is from Gallagher et~al. (1999).}
\tablenotetext{c}{This is a steep spectrum radio-loud QSO. All the others are radio-quiet QSOs.}
\tablenotetext{d}{This \aox\ value is from Elvis \& Fabbiano (1984).}
\end{deluxetable}

\clearpage


\begin{deluxetable}{lll}
\tablenum{3}
\tablewidth{0pt}
\tablecaption {Kolmogorov-Smirnov (KS) and Kuiper test probabilities comparing the 
continuum and optical emission-line properties of SXW~QSOs and non-SXW~QSOs}
\small
\tablehead{
\colhead{Variable name$^{\rm a}$}   & 
\colhead{KS test probability}       & 
\colhead{Kuiper test probability}             
}
\startdata
$M_{\rm V}$               & 0.4113         & 0.1671      \nl
$M_{\rm [O~{\sc iii}]}$   & 0.0156         & 0.0091      \nl
$z$                       & 0.9492         & 0.7500      \nl
$\log R$                  & 0.3353         & 0.8108      \nl
EW H$\beta$               & 0.7051         & 0.5711      \nl
EW~$\lambda 5007$         & 0.0324         & 0.2013      \nl
EW~$\lambda 4686$         & 0.1856         & 0.6153      \nl
EW~Fe~{\sc ii}            & 0.1823         & 0.5901      \nl
[O~{\sc iii}]/H$\beta$    & 0.0959         & 0.4425      \nl
He~{\sc ii}/H$\beta$      & 0.1856         & 0.2886      \nl
Fe~{\sc ii}/H$\beta$      & 0.2694         & 0.6717      \nl
FWHM H$\beta$             & 0.4902         & 0.6779      \nl
Shift H$\beta$            & 0.1205 (80)    & 0.0032 (80) \nl
Shape H$\beta$            & 0.0017         & 0.0122      \nl
Asymmetry H$\beta$        & 0.1237         & 0.5211      \nl
%
%
$\alpha_{\rm opt}$        & 0.7998 (74)    & 0.7442 (74) \nl 
Eigenvector~1             & 0.0012         & 0.0062      \nl
Eigenvector~2             & 0.1190         & 0.5088      \nl
`\aox-free' eigenvector~1 & 0.0121         & 0.0195      \nl
%
%
%
\enddata
\tablenotetext{a}{See BG92 for detailed explanations of these 
variable names.}
\tablenotetext{b}{Unless noted otherwise, all tests were performed
using the 87 BG92 QSOs. When quantities were not available 
for all 87 BG92 QSOs, we list the number that were used in 
parentheses after the relevant probability.}
\end{deluxetable}

                                                                        

\begin{deluxetable}{llc}
\tablenum{4}
\tablewidth{0pt}
\tablecaption {Spearman rank-order correlations against \aox\ when our SXW~QSOs are removed}
\small
\tablehead{
\colhead{Variable name$^{\rm a}$}           & 
\colhead{Spearman $R^{\rm b}$}              & 
\colhead{Spearman probability}             
}
\startdata
$M_{\rm V}$               &   $+0.2268$       &  0.0472  \nl
$M_{\rm [O~{\sc iii}]}$   &   $-0.0268$       &  0.8168  \nl
$z$                       &   $-0.1411$       &  0.2207  \nl
$\log R$                  &   $+0.0915$       &  0.4287  \nl
EW H$\beta$               &   $+0.2653$       &  0.0197  \nl
EW~$\lambda 5007$         &   $+0.2525$       &  0.0267  \nl
EW~$\lambda 4686$         &   $+0.2884$       &  0.0109  \nl
EW~Fe~{\sc ii}            &   $-0.2968$       &  0.0088  \nl
[O~{\sc iii}]/H$\beta$    &   $+0.1424$       &  0.2167  \nl
He~{\sc ii}/H$\beta$      &   $+0.2380$       &  0.0371  \nl
Fe~{\sc ii}/H$\beta$      &   $-0.3838$       &  0.0006  \nl
FWHM H$\beta$             &   $+0.2392$       &  0.0361  \nl
Shift H$\beta$            &   $-0.0806$ (72)  &  0.5006  \nl
Shape H$\beta$            &   $-0.1638$       &  0.1545  \nl
Asymmetry H$\beta$        &   $-0.3503$       &  0.0018  \nl
%
%
$\alpha_{\rm opt}$        &   $-0.2954$ (65)  &  0.0169  \nl 
Eigenvector~1             &   $+0.2917$       &  0.0100  \nl
Eigenvector~2             &   $+0.3077$       &  0.0065  \nl
`\aox-free' eigenvector~1 &   $+0.2408$       &  0.0349  \nl
%
%
%
\enddata
\tablenotetext{a}{See BG92 for detailed explanations of these 
variable names.}
\tablenotetext{b}{Unless noted otherwise, all Spearman $R$ values were 
computed using the 77 non-SXW~QSOs. When quantities were not available 
for all 77 non-SXW~QSOs, we list the number that were used in 
parentheses after the Spearman $R$ value.}
\end{deluxetable}

                                                                        

\begin{figure}
\epsscale{0.5}
\plotfiddle{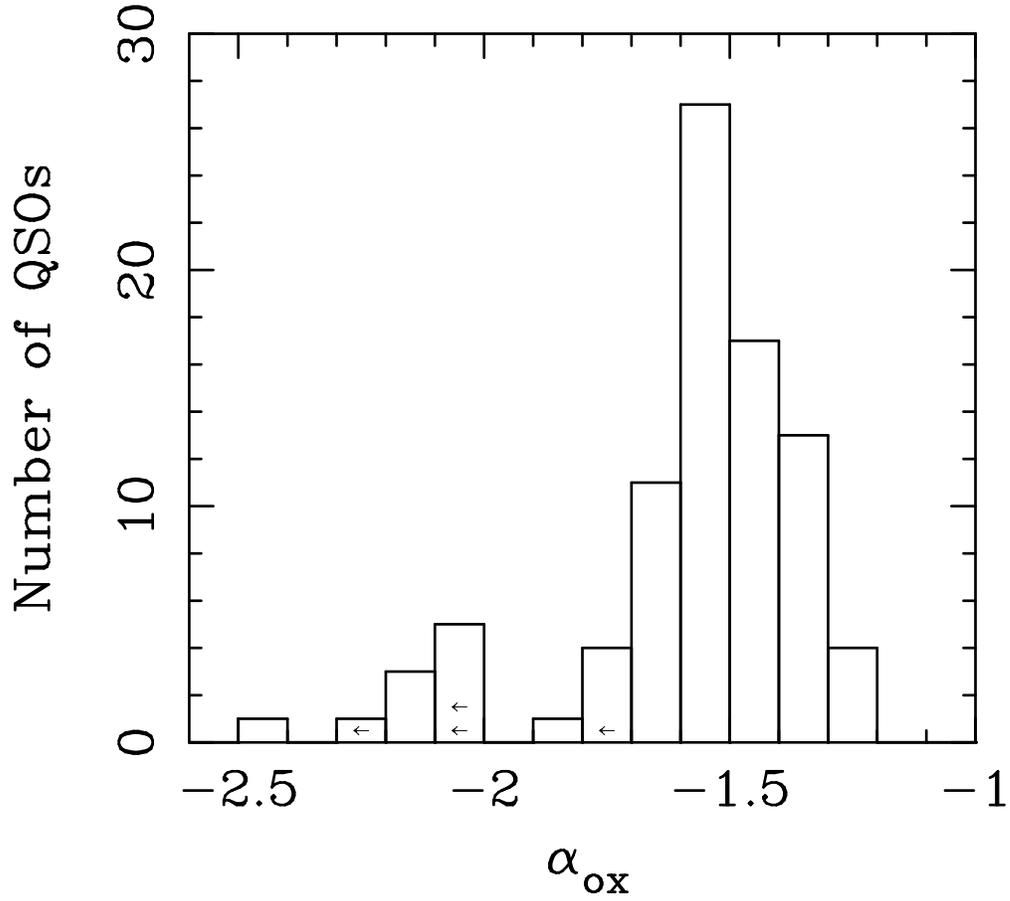}{360pt}{-90}{70}{70}{-240}{+500}
\vspace{-1.5in}
\caption{Histogram showing the \aox\ distribution of the 87 BG92 QSOs. 
The values of \aox\ refer to the positions indicated by the longer 
tick marks on the top axis. The four nondetections present in this 
histogram (0043+039, 1004+130, 1259+593 and 1700+518) are labeled with 
left-pointing arrows. 
\label{fig1}}
\end{figure}
 


\begin{figure}
\epsscale{0.5}
\plotfiddle{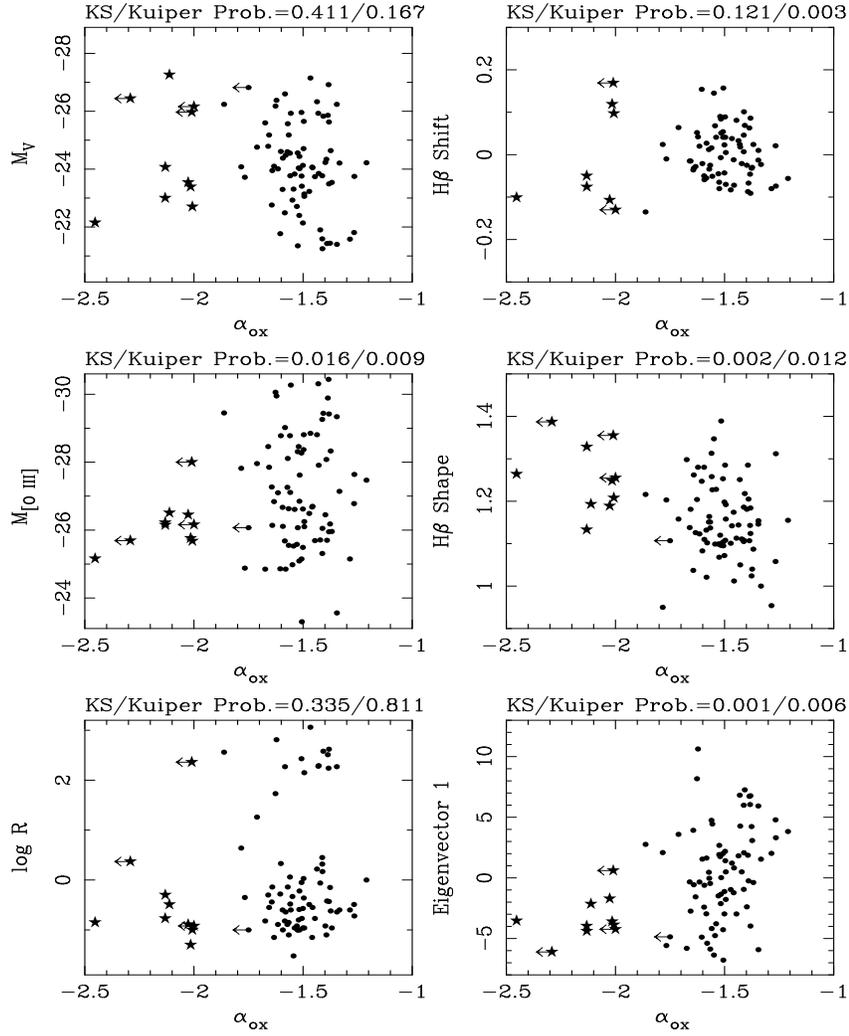}{360pt}{0}{70}{70}{-220}{-40}
\vspace{-0.6in}
\caption{Comparisons of the continuum and optical emission-line properties
of our 10 SXW~QSOs (stars, referred to as SXW in the text) to the other 77 
BG92 QSOs (dots, referred to as non-SXW in the text). We do not
show plots for all the properties discussed in the text but rather 
focus on the most germane ones. Above each panel are listed the 
Kolmogorov-Smirnov (KS) and Kuiper test probabilities that the stars
and dots are drawn from the same distribution. Note that the 
two samples are consistent with being drawn from the same $M_{\rm V}$ and 
$\log R$ distributions but that the SXW~QSOs differ in terms of the 
other illustrated properties. 
\label{fig2}}
\end{figure}
 
 

\begin{figure}
\epsscale{0.5}
\plotfiddle{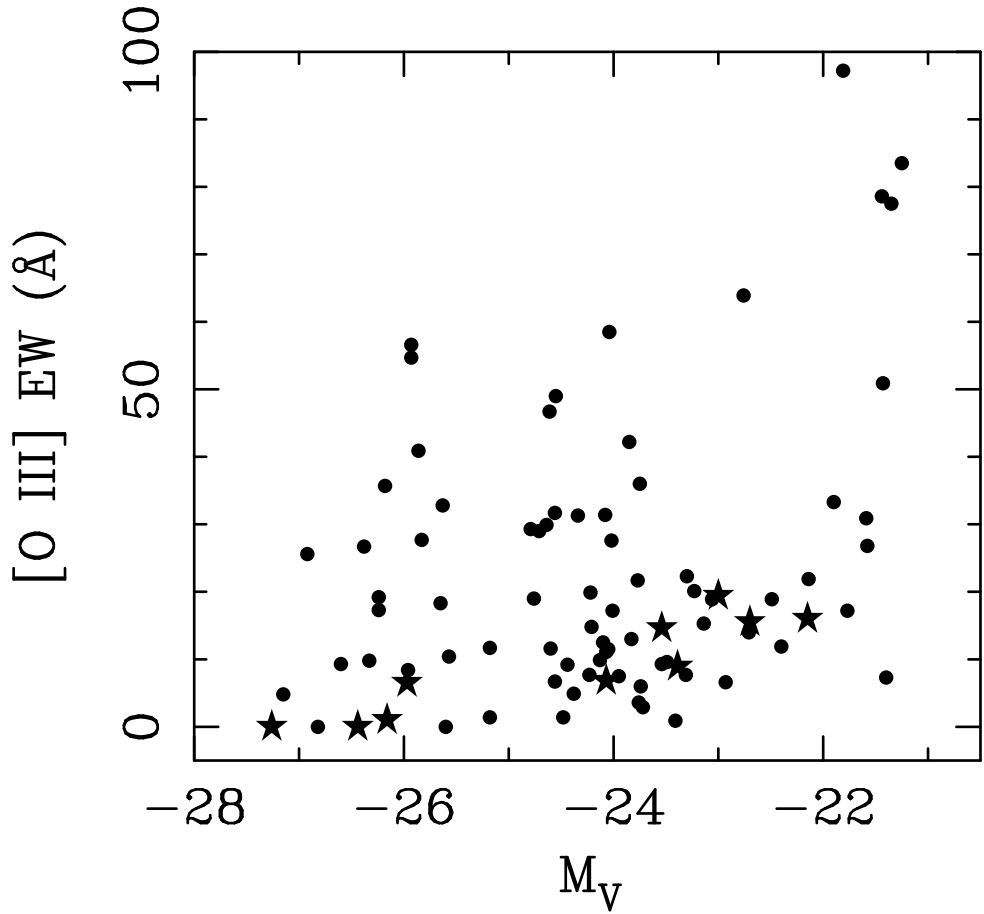}{360pt}{-90}{70}{70}{-240}{+500}
\vspace{-0.6in}
\caption{Plot of [O~{\sc iii}] EW versus $M_{\rm V}$ for the
BG92 QSOs. Stars show our 10 SXW~QSOs, and dots show the 
other BG92 QSOs. Note that the SXW~QSOs have systematically
low [O~{\sc iii}] EWs. The point for 1612+261 lies outside
the boundary of this plot at $M_{\rm V}=-23.77$, 
[O~{\sc iii}]~EW~$=157$~\AA. 
\label{fig3}}
\end{figure}
 
 

\begin{figure}
\epsscale{0.5}
\plotfiddle{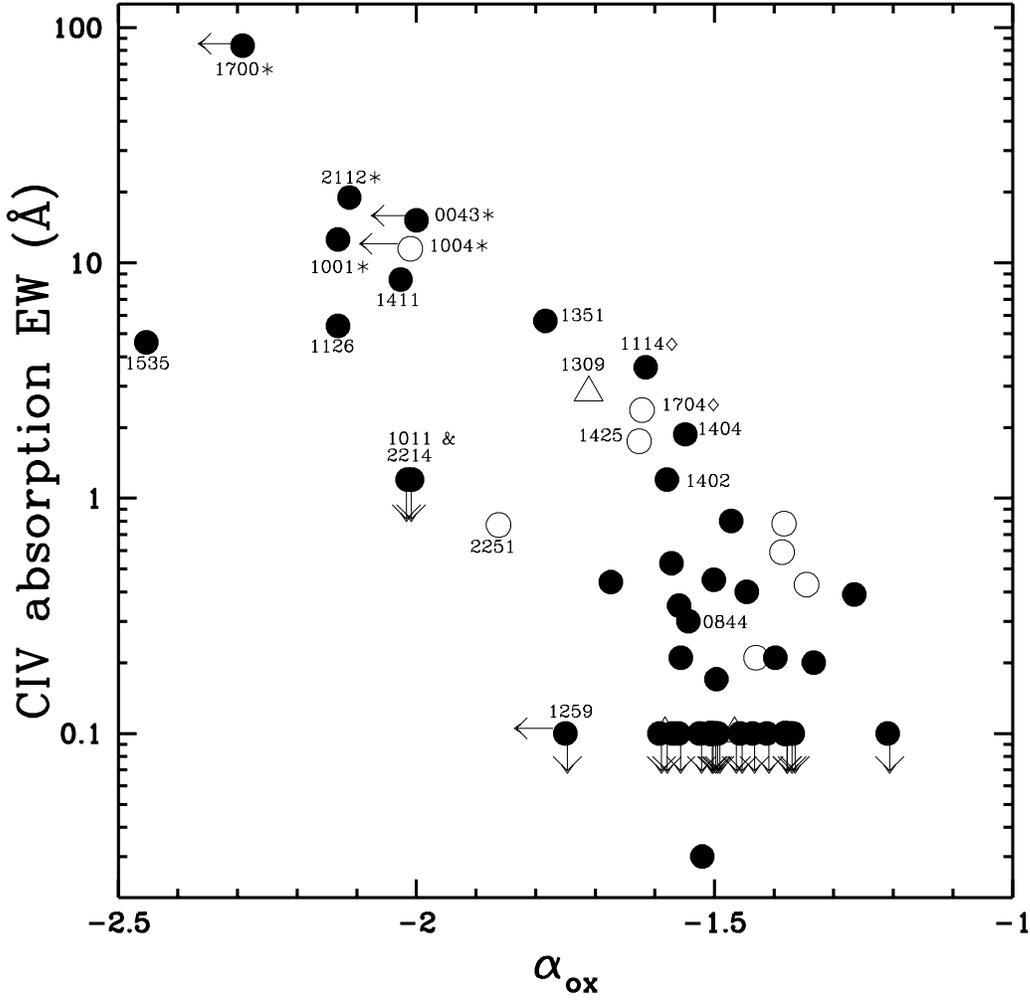}{360pt}{0}{70}{70}{-220}{10}
\vspace{-1.4in}
\caption{Plot of \aox\ versus C~{\sc iv} absorption-line EW 
(in the rest frame) for the BG92 QSOs. Following BG92, solid dots are 
radio-quiet QSOs, open triangles are core-dominated radio-loud QSOs,
and open circles are lobe-dominated radio-loud QSOs. SXW~QSOs and
other particularly relevant objects are labeled by the right ascension 
part of their BQS name. An asterisk ($\ast$) to the right of an object's 
name indicates that it is a BAL~QSO or probable BAL~QSO
(see \S5 for further discussion), and a diamond ($\diamondsuit$) to
the right of an object's name indicates that it is known to have
an X-ray warm absorber. We have assigned an EW of $<0.1$~\AA\ 
when no C~{\sc iv} absorption was detected. 
\label{fig4}}
\end{figure}
 
 

\begin{figure}
\epsscale{0.5}
\plotfiddle{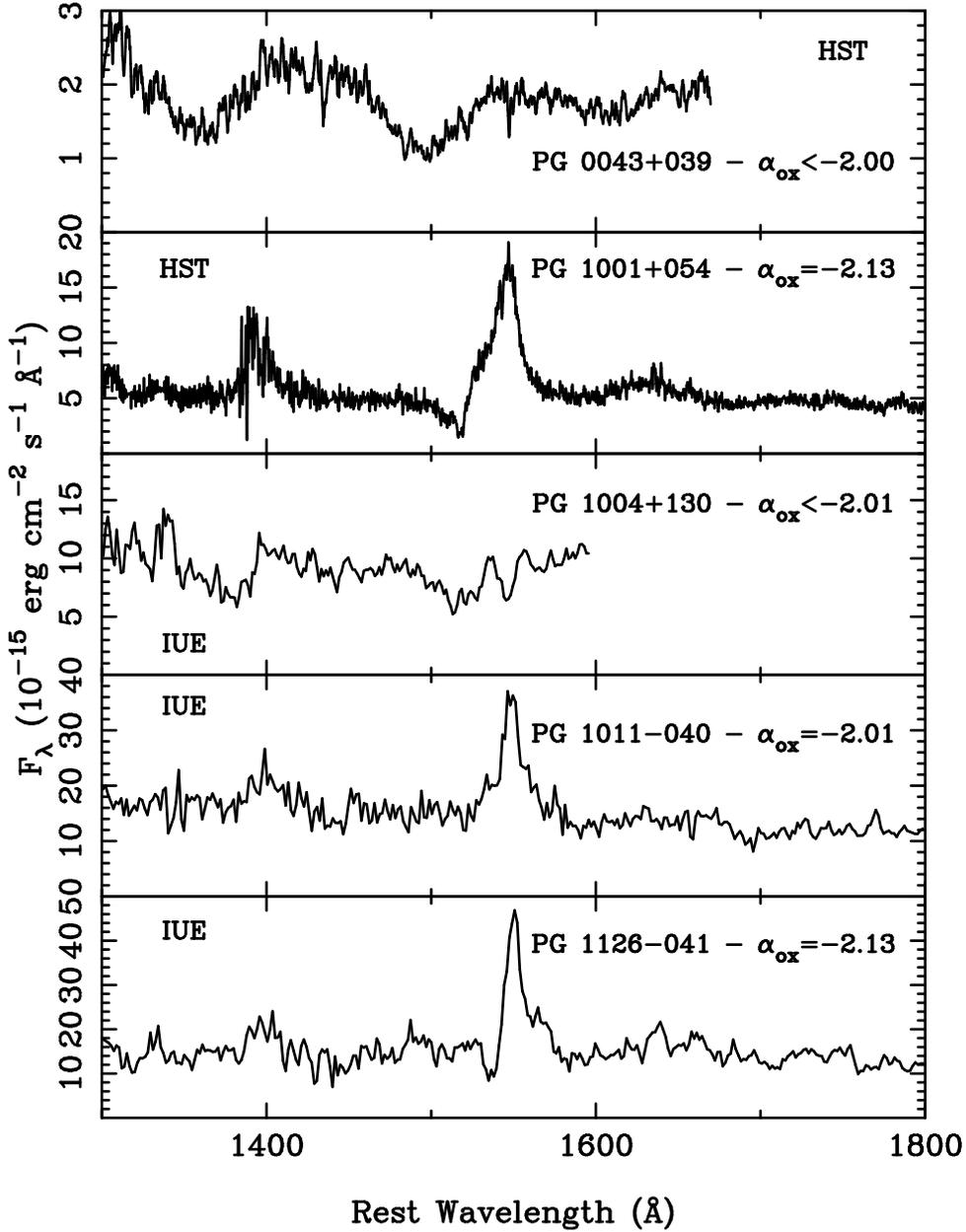}{360pt}{0}{70}{70}{-220}{-40}
\vspace{0.5in}
\caption{Spectra of the regions of the broad C~{\sc iv} emission line for
our ten SXW~QSOs. Note the strong, blueshifted C~{\sc iv} absorption seen 
in most spectra. In each panel we also give the \aox\ value and the name 
of the satellite that took the data shown. 
\label{fig5a}}
\end{figure}
 
 


\begin{figure}
\epsscale{0.5}
\plotfiddle{fig5b.ps}{360pt}{0}{70}{70}{-220}{-40}
\vspace{0.5in}
\end{figure}
 
 

\begin{figure}
\epsscale{0.5}
\plotfiddle{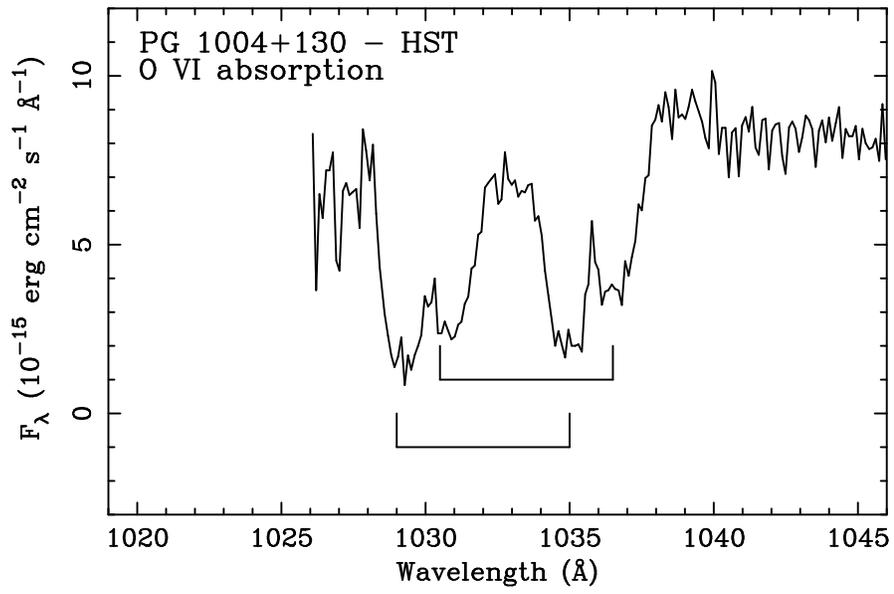}{360pt}{-90}{50}{50}{-200}{420}
%
\vspace{-1.5in}
\caption{Archival \hst\ GHRS spectrum showing the strong O~{\sc vi}
absorption present in the ultraviolet spectrum of 1004+130.
Two absorption systems are seen at $z=0.2364$ and $z=0.2387$. 
%
\label{fig6}}
\end{figure}

\clearpage


\end{document}